# Proposing a Dynamic Executive Microservices Architecture Model for AI Systems

Mahyar Karimi
Amirkabir University of Technology (Tehran Polytechnic)
Tehran, Iran
mahyarkarimi@aut.ac.ir
Tehran, Iran

Prof. Ahmad Abdollahzadeh Barforoush
Amirkabir University of Technology (Tehran Polytechnic)
ahmadaku@aut.ac.ir
Tehran, Iran

*Abstract*— Microservices architecture is one of the new architectural styles that has improved in recent years. It has become a popular architectural style among system architects and developers. This popularity increased with the advent of new technologies and technological advancements in cloud computing. These advancements caused the emergence of new design and development challenges for service-based software systems. The increasing use of microservices architecture in large organizations and teams has increased the need to find appropriate solutions for architecture challenges. Orchestration of the components in the microservices architecture is one of the main challenges in distributed systems and affects the software quality in factors such as efficiency, compatibility, stability, and reusability. In such systems, software architecture consists of fine-grained components. Due to the increasing number of microservices in a large-scale system, proper management and communication orchestration of microservice components can become a point of failure. In this article, the challenges of Microservices architecture have been identified. To resolve the component orchestration challenges, an appropriate model to maintain and improve quality is proposed. The presented model, as a pattern, can be used at the both design and development level of the system. The Dynamicity of software at runtime is the main achievement of this pattern. In this model, microservice components orchestration tasks are performed by using a BPMN-based workflow engine as the orchestrator component. The orchestrator design gives the ability to create, track and modify new composite microservices without the need to change platform infrastructure.

## I. Introduction

Using the Microservices architecture in the design and implementation of software systems leads to achieving better quality by changing the software development lifecycle [1][2]. The main approach to solving architectural challenges in service-oriented architecture is to divide software into smaller components. This approach originates from the separation of concerns and R.C. Martin's single repository principles [17][18]. As a result of dividing software into smaller components quality factors like maintainability and performance increase [5]. Implementing software as a set of fine-grained microservices components leads to a reduction in maintenance costs [2]. This also removes the need to create Single Viewpoint models of stakeholder requirements, which improves the dynamicity and simplicity.

Considering continuous integration and continuous delivery alongside architectures consisting of separate components can conclude that the microservices architecture is an appropriate solution for distributed software systems and clouds [9][13][14].

In this article, a design pattern or more generally an approach consisting of instructions and principles is proposed with the goal to improve the dynamicity of AI systems with the microservices architecture.

## II. Architecture Framework

Fundamental concepts or properties of a system in its environment are embodied in its elements, relationships, and its principles of its design and evolution [4]. The environment which influences the system is defined as settings and circumstances determined by context [4]. Software architectures are mainly categorized into eight groups [6]:

1. Distributed
2. Monolithic
3. Pipes & Filter
4. Layered
5. Event-Driven
6. Peer-to-Peer
7. Blackboard
8. Data-Centric.

These architectures and the factors related to the microservices architecture are shown in Fig 1, which implies the taxonomy of the software architecture.

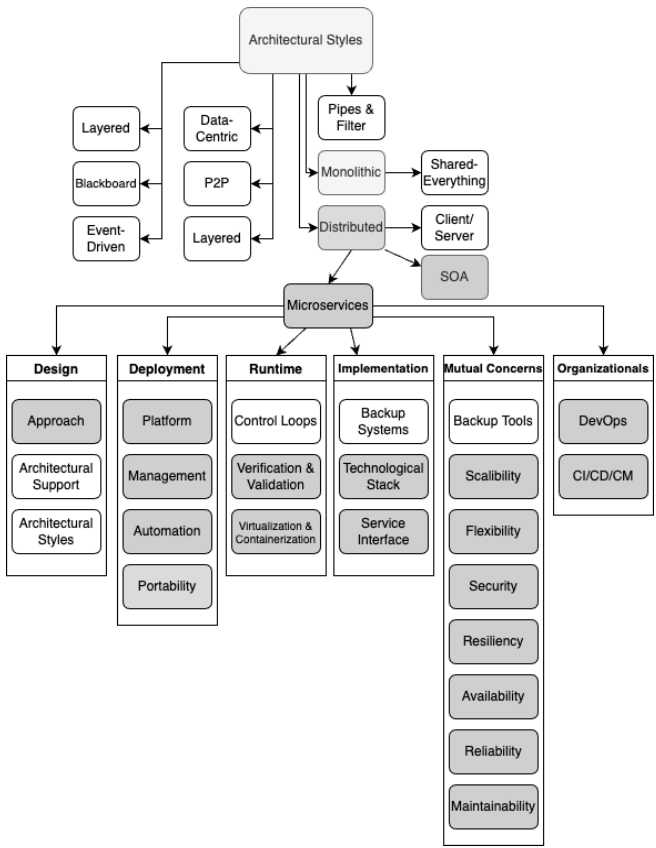

*Figure 1 – Taxonomy of software architecture and principles covered in this article.*

According to the taxonomy of software architectures, factors and principles of the distributed architectures and specifically the Microservices architecture are examined in the following section. Di Francesco et al. denote that most of the articles focus on proposing a solution to overcome complexity, flexibility, and performance challenges in the Microservices architecture-based systems [15].

Considering architecture framework as a set of principles and high-level instructions that are used for a specific scope of systems with specified functionalities. In order to generalize the framework to be applicable to any system with the same specification, the principles and instructions should be repeatable for different contexts.

*A. Service-Oriented & Microservices Architecture*

The Service-Oriented architecture is a distributed software architecture based on the service-oriented computing paradigm [12]. Components of architecture can be distributed geographically or virtually. Service-Orientation originates from OOP where different components of software programs communicate via lightweight network protocols [1][2][5]. From a software architecture point of view, service-oriented architecture is defined as a set of interconnected services [16]. The term Microservice was first introduced in a software architecture workshop in Venice in 2011 [2]. Focused on implementing functionalities as simple and small components, the Microservices architectural style is derived from service orientation in order to reduce unnecessary complexity [2][12]. This also results in coupling reduction between microservices and cohesion increase inside each microservice. In the microservices architecture, one of the goals is to distribute software components in a way that each component can reach high levels of autonomy and independence while interacting with other parts of the system. Enhancement of software quality in aforementioned factors like complexity and reusability, the software development lifecycle alongside operational tasks gains a significant acceleration [5][7][15][16][21].

Table 1 shows a comparison between the traditional monolithic, service-oriented, and microservices architectures for principles and quality criteria. As a result of this benchmark, the pros of the microservices architecture are noticeable.

### III. Effects of the Microservices architecture on requirements engineering

Using the microservices architecture in software design and implementation has an impact on both architecturally significant requirements and user requirements [23]. User requirements will meet more dynamically with lower runtime costs. To achieve dynamicity, the architecture of software must follow principles of the microservices architecture which may change the design and add crucial components to it. Based on the proposed architectural style, with trying to maximize the reusability and composability of components in horizontal and vertical dimensions, we lay claim to having achieved a system that can meet any user requirement with much lower runtime cost and time. The service composition approach is the key concept in the proposed model [1][2][15][17]. Although there is a significant increase in costs which refers as compile time cost, as a result of its much lower runtime costs mentioned earlier, it is preferable. Also, there is no need for software engineers to carry out requirement single view process in model-based requirements engineering [23]. Having the ability to provide dynamic and multiple microservice version support platforms arises no need for a single view process and outcomes in a more user-customized environment. The method of developing software with consideration of multiple stakeholders' viewpoints requires resolving conflicts of requirements and adjusting a single viewed SRS document. This is a rather expensive and sometimes hard to perform requirements engineering of the software lifecycle.

Architectural significant requirements, that are essential to the architecture, are derived from the microservices architecture design principles. The proposed design model covers ASR as well as the other instructions needed to apply the pattern.

### IV. Targeted challenge and previous work

The proposed solution for AI systems from an IT perspective is the concept of service-oriented architectures and microservices. While a microservices architecture addresses challenges of producing enterprises, new challenges come into view by their implementation. Among these challenges is the architecture design in terms of communication of the microservices or their orchestration and integration [1]. One of the key challenges in decomposing large services into fine-grained microservices is the communication of smaller

| Pattern | Origin | Goal | Features | Pros & Cons | Used in |
|---|---|---|---|---|---|
| API Gateway | Based on SOA principles | Improve simplicity | Better encapsulation, security, abstraction, routing, transformation, and other policies | Pros: simpler development and backward compatibility, market-based [3][32][53]<br>Cons: possible single point of failure, complex for a large number of services [32][58] | [3][11][29][30][31][32][33][34][35][83][39] |
| Client-side Discovery | Developed by Netflix and Pivotal [29] | Find relative service dynamically | Clients have the ability to choose a service. load balancing can be applied through this pattern, direct connection between consumer and service | Pros: simple usage<br>Cons: increase coupling between consumers and selecting services [12] | As reported by Richardson in [11][29] |
| Server-side Discovery | | Provide direct connection to services with a central namespace (like a DNS provider) | Direct connection between consumer and service after client resolved service address through the registry, can be used as DNS server. | Pros: straightforward maintenance and provide the ability to use names for services. Better understandability and much simpler logic compared to client-side discovery [1][11][29][38].<br>Cons: service interface changes may break functionality, possible single point of failure | [39][40][42][43][44][45][46][47][55] |
| Hybrid API-Gateway | | Replace service bus with a combination of API-Gateway and Registry | Communications between client and microservices, and inter microservices can be conducted with this component. The gateway is responsible for routing the targeted microservices. | Pros: Migration of system architecture with this pattern does not affect current clients and can be achieved gradually. Faster learning curve [3]. | [48][49][50][51][52][53] |
| Multiple-instance-per-host | Originates from deployment principle of microservices architecture | Improve scalability of microservices | This pattern is the opposite of the Single-instance-per-host pattern. The main advantage of the single instance pattern is this approach of complete isolation, however, this dramatically reduces performance and scalability, and dedicating a single node to a microservice violates the basic idea of microservices. | Pros: Efficient use of resources, simpler scalability of services.<br>Cons: Less isolation between the same instances on the host. Monitoring resource usage of services has more effort. | [30][35][38][39][42][45][50][51][52][53][54][55][56][57][58][59] |

components. Cooperation of independent microservices in order to provide a bigger service can be accomplished with orchestration or choreography [1][2][5]. Choreography is the approach of communication managed by components themselves while orchestration is a more centric approach that requires an orchestrator [].

Generally, architecture design patterns play a significant role in a microservices architecture. These patterns are grouped into three main categories [11]:

1. Orchestrations and coordination patterns
2. Deployment patterns
3. Data storage patterns

We focus on orchestration and coordination-oriented architecture patterns that capture communication and coordination of components from a logical point of view [2]. While there are several patterns targeting orchestration and coordination criteria, some of the most used and popular patterns are discussed in the following table in the OAV[1] model.

V. MODELING THE ARCHITECTURAL STYLE PATTERN

In order to have a deeper understanding of the targeted challenge and scope of the proposed solution, the problem and solution domains are specified according to the requirements engineering generic model from E. Hull Requirement Engineering [23].

---

[1] Object Attribute Value

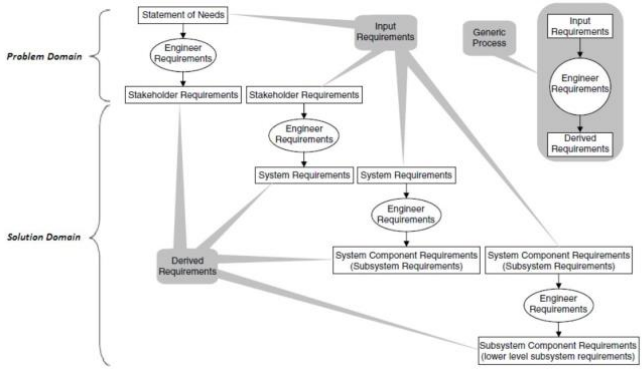

*Figure 2 - Requirements Engineering Process Model [23]*

In previous sections, we have discussed the influence of the proposed model on requirements engineering and the single view process, henceforth the user and architectural significant requirements are listed to address the need for achieving a higher quality product in cloud and microservice-based environments.

### A. The Problem and Solution Domain

According to Hull etc., all [23] problem domain is defined as the domain in which a system is going to be used. This is the heart of requirements engineering in which the elicitation of requirements is done by asking questions like `What the system must be able to do? ` [23]. Using Boehm's W5HH principle [] the objectives, timelines, responsibilities, management styles, and resources are specified. For the context of the proposed orchestration pattern these W5HH-like questions are listed as follows:

- What is the purpose of the design and development of this pattern?
- Why the pattern is developed?
- When this pattern can be applied?
- Who can use this pattern in the software development lifecycle?
- Which software types can be designed and developed using this pattern?
- How will the application of this pattern be done technically?
- How much quality gain will be provided by the application of this pattern?

With a given set of requirements, the problem domain does not solve the problem. Although, in the problem domain the requirements are well-formed alongside the elicitation of the Architectural Significant Requirements [23].

As it is shown in Fig 1, modeling and analysis are done first to provide a basis for deriving the requirements secondly [23]. Half of the system pattern is defined from the scope, and requirements from the Fig 1 process model. The properties, relationships of elements, and its design modeling will be described.

### B. BPMN Standard

The BPMI[2] has developed a standard Business Process Modeling Notation (BPMN) which has a primary goal of providing a notation that has a high quality of understandability by all business users, from analysts to technical developers and operation managers [10][19][20][22][27][28]. The BPMN creates a standardized bridge to fill the gap between design and implementation in the process [10]. It is based on a flowcharting technique consisting of a network of graphical objects, which are activities and the flow controls [10][19][29]. BPMN as a well-supported standard modeling language reduces the confusion among different stakeholders [10].

Use case descriptions of complex procedures are rather difficult to understand and prone to different errors. Since a clear picture depicting a workflow is in most cases self-explaining, many users aim to enrich descriptions of processes with diagrams in order to convey the intended meaning associated with the process. Moreover, examining a graphical description of a process allows users to easily discover inconsistencies, differences in names, infinite loops, non-terminating conditions, and in general, the flow of the procedure to be done [19].

By using a complete de facto standard like BPMN, the link between analysis, design, and implementation becomes robust and resilient to defects. With the use of BPMN, we are going to design a microservices orchestration engine, which orchestrates based on workflows defined as BPMN standards potentially independent of the infrastructure of deployment, although the implementation assumes a container-based deployment in a Kubernetes cluster.

### C. Scope of the Pattern & Achieved Requirements

For an architectural pattern based on a specific architectural style, there exist multiple aspects. One of these aspects is the base principles forming the baseline components and their relations. Also, the quality model, scope, and requirements are the other aspects. The quality model and the base principles are discussed earlier. Table 2 and 3 defines the scope and requirements.

To distinguish scope in the context of the proposed pattern, the collection of repeatable and coordinated tasks in order to fulfill a process is defined as a workflow [24][25]. In a more abstract view, workflow is a representation of the process to be done [26].

*Table 2 - Scope of Workflow Engine Orchestration Pattern*

| Scope | Coverage Status |
|---|---|
| Prefer the microservices orchestration approach over choreography | ✓ |
| Propose a pattern with the capability of serverless computing. | ✓ |
| Generalize the architectural pattern to support different contexts and software types. | ✓ |

---

[2] Business Process Management Initiative

| Container orchestration | ✗ |
|---|---|
| Microservices architecture-specific quality model | ✗ |

The following table consists of user and architecture requirements addressed to meet by the application of this pattern.

*Table 3 – Problem & Solution Domain Requirements*

| User Requirements | |
|---|---|
| Description | Q Factor |
| Developers must not have concerns related to scalability, test, and deployment of their microservices in the process of development. | Usability, Maintainability |
| The system should have the ability to create new services from composing previous microservices. | Reusability |
| The system should support multiple service versioning with the lowest effort. | Compatibility, Functionality |
| Developers should be able to create services by the creation of a chain of microservices with no concern for service input/output communication. | Functionality, Reusability, Interoperability |
| Developers should be able to choose microservices from a repository pool of microservices in the process of composite service creation. | Portability, Reusability, Compatibility |
| Architectural Significant Requirements | |
| Description | Q Factor |
| Conformity of BPMN2 standard to the execution and data flow model of microservices orchestration | Functional, Performance |
| Design and development of workflow engine compatible with known most used container orchestration platforms like Kubernetes. | Functional, Compatibility, Interoperability |
| The workflow engine must be able to read and parse the execution flow based on standard visualized notation language BPMN2. | Functional |
| The workflow engine should provide a graphical user interface to view, create, edit and remove workflows. | Functional, Maintainability |
| The GUI of the workflow engine must have a BPMN2 editor. | Functionality |
| FaaS-based SDKs can be developed or using an existing one to support serverless computing to provide more simplicity and efficiency. | Functionality, Reusability, Usability |

*D. Modeling & Design*

Fig 3 illustrates a context diagram of a system designed by application of workflow engine orchestration pattern. Four main tasks are comprehended which consist of management of BPMN-based workflows and microservices, creation of microservices from source in a serverless computing model, monitor the state of workflows and microservices. With the separation of these functionalities into an external independent component, the development of the system will become simpler and more microservice-like. These tasks are not related to microservice development and therefore must be dealt with in another scope. By doing so, the focus of a developer will be on the main functionality of a microservice which leads to maximization of quality.

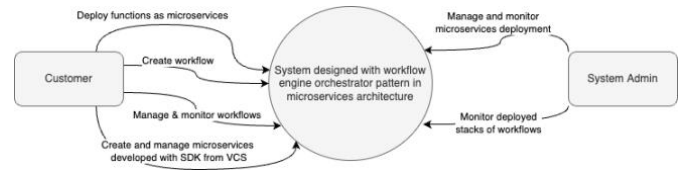

*Figure 3 - System Context Diagram Design with Workflow Engine Pattern*

To gain deeper insight into how the orchestration is done by the workflow engine, Fig 4 shows the flow of data and interoperable events between different elements of system architecture. The elements in the DFD diagram are considered to behave like components, which means they can be replaced by their alternative.

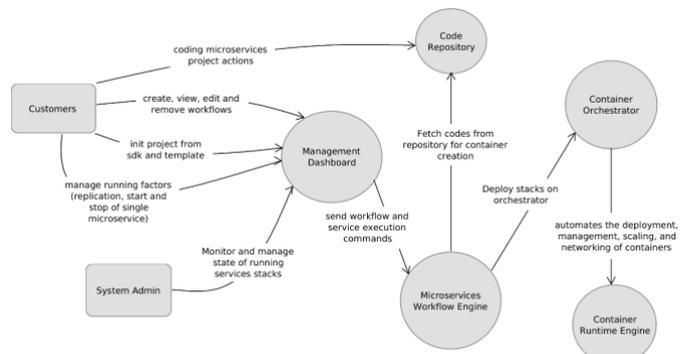

*Figure 4 - System Data Flow Diagram Orchestrated Using Workflow Engine*

At first sight, it is obvious that a workflow engine orchestrator is not a container orchestrator and does not have provision over microservices resources and their deployment, however, it uses the container orchestrator to perform some of its jobs. The model explicitly specifies that every customer can create service either by deploying a single microservice or a workflow based on the BPMN document which is considered as a composition of microservices. The orchestrator engine has the ability to parse the BPMN document and perform the execution flow of each microservice defined in a document in exact order.

The BPMN version 2.0 supports different execution logic flows, which the orchestrator should be able to execute according to the standard. Also, many other characteristics and functionalities can be handled within the workflow engine including circuit breaking, retry policies, high availability policies, fault tolerance, etc.

## VI. Conclusions & Evaluation

To summarize the specifications, it can be inferred that the 'Workflow Orchestration Pattern' originates from a traditional BPMN engine combined with an API-Gateway pattern adapted into a modern containerized environment. It empowers software with a centralized management element based on BPMN and a higher reusability level alongside all the benefits of the API-Gateway pattern and the BPMN standard.

The main approach of this pattern is to overcome communication challenges that emerged in microservice-based systems. A survey of previous works shows that communication is one of the essential challenges.

## Acknowledgment (Heading 5)

The preferred spelling of the word "acknowledgment" in America is without an "e" after the "g." Avoid the stilted expression "one of us (R. B. G.) thanks ...". Instead, try "R. B. G. thanks...". Put sponsor acknowledgments in the unnumbered footnote on the first page.

## References

[1] J. Lewis and M. Fowler, "Microservices", https://martinfowler.com/articles/microservices.html, 2014.

[2] S. G. Nicola Dragoni, Alberto Lluch Lafuente, Manuel Mazzara, Fabrizio Montesi, Ruslan Mustafin, Larisa Safina, "Microservices: Yesterday, Today, and Tomorrow," in Present and Ulterior Software Engineering, B. M. Manuel Mazzara, Ed.: Springer, Cham, 2017, pp. 195-216.

[3] D. Namiot und M. Sneps-Sneppe, "On micro-services architecture," *International Journal of Open Information Technologies*, p. 24-27, 2014.

[4] "Systems and software engineering Architecture description", ISO/IEC/IEEE 42010:2011(E), 2011.

[5] S. Newman, *Building microservices: designing fine-grained systems*, 2. Ed., London: " O'Reilly Media, Inc.", 2015.

[6] M. Garriga, "Towards a taxonomy of microservices architectures," in International Conference on Software Engineering and Formal Methods, 2017: Springer, pp. 203-218.

[7] T. Erl, Service-Oriented Architecture: Concepts, Technology, and Design. Pearson Education, 2005.

[8] "Chapter 1: Service Oriented Architecture (SOA)". msdn.microsoft.com. Archived from the original on February 6, 2016. Retrieved September 21, 2016.

[9] A. Balalaie, A. Heydarnoori, and P. Jamshidi, "Microservices Architecture Enables DevOps: Migration to a Cloud-Native Architecture," IEEE Software, vol. 33, no. 3, pp. 42-52, 18 March 2016 2016.

[10] S. A. White, 'Introduction to BPMN', *IBM Cooperation*, 2004.

[11] C. Richardson, "Microservices patterns," ed: Manning Publications, Shelter Island, 2018.

[12] F. Rademacher, S. Sachweh, and A. Zündorf, "Differences between Model-Driven Development of Service-Oriented and Microservice Architecture," in 2017 IEEE International Conference on Software Architecture Workshops (ICSAW), 2017, pp. 38-45.

[13] J. Humble and D. Farley, Continuous Delivery: Reliable Software Releases through Build, Test, and Deployment Automation (Adobe Reader). Pearson Education, 2010.

[14] L. J. I. S. Chen, "Continuous delivery: Huge benefits, but challenges too," vol. 32, no. 2, pp. 50-54, 2015.

[15] P. D. Francesco, P. Lago, and I. Malavolta, "Architecting with microservices: A systematic mapping study," Journal of Systems and Software, vol. 150, pp. 77-97, 2 January 2019 2019.

[16] D. K. Barry und D. Dick, „Chapter 3 - Web Services and Service-Oriented Architectures," Second Edition. Ed., in Web Services, Service-Oriented Architectures, and Cloud Computing (Second Edition), Boston, Morgan Kaufmann, 2013, p. 15 - 33.

[17] T. Erl, *SOA Principles of Service Design*. Pearson Education, 2007.

[18] R. C. Martin, Agile software development: principles, patterns, and practices. Prentice Hall, 2002.

[19] M. Chinosi και A. Trombetta, 'BPMN: An introduction to the standard', *Computer Standards & Interfaces*, τ. 34, τχ. 1, σσ. 124–134, 2012.

[20] R. C. Simpson, "An XML representation for crew procedures", *NASA Summer Faculty Fellowship Program 2004,* Volumes 1 and 2, 2005.

[21] T. Erl, *SOA Principles of Service Design*. Pearson Education, 2007.

[22] S. A. White en Others, "Process modeling notations and workflow patterns", *Workflow handbook*, vol 2004, bll 265–294, 2004.

[23] E. Hull, J. Dick and K. Jackson, *Requirements Engineering*, 4th ed. Springer International PU, 2018.

[24] Business Process Management Center of Excellence Glossary, 2009. Accessed On Nov 10, 2021. [Online] Available: https://www.ftb.ca.gov/aboutFTB/Projects/ITSP/BPM_Glossary.pdf

[25] M. Hammer, "What is business process management?", in Handbook on business process management 1, Springer, 2015, bll 3–16.

[26] ISO Standard Health informatics - Digital imaging and communication in medicine (DICOM) including workflow and data management, ISO 12052:2017

[27] Stephen A. White. "Business Process Modeling Notation v1.0" (PDF). Archived from the original (PDF) on 18 August 2013. for the Business Process Management Initiative (BPMI).

[28] Omg.org. 2021.About the Business Process Model And Notation Specification Version 2.0. Accessed On Nov 11, 2021. [online] Available at: https://www.omg.org/spec/BPMN/2.0/

[29] Richardson, C, "Microservice Architecture," http://microservices.io.

[30] M. Villamizar et al., "Evaluating the monolithic and the microservice architecture pattern to deploy web applications in the cloud", in 2015 10th Computing Colombian Conference (10CCC), 2015, bll 583–590.

[31] S. Alpers, C. Becker, A. Oberweis, en T. Schuster, "Microservice based tool support for business process modelling", in 2015 IEEE 19th International Enterprise Distributed Object Computing Workshop, 2015, bll 71–78.

[32] D. Malavalli en S. Sathappan, "Scalable microservice based architecture for enabling dmtf profiles", in 2015 11th International Conference on Network and Service Management (CNSM), 2015, bll 428–432.

[33] P. Bak, R. Melamed, D. Moshkovich, Y. Nardi, H. Ship, en A. Yaeli, "Location and context-based microservices for mobile and internet of things workloads", in 2015 IEEE International Conference on Mobile Services, 2015, bll 1–8.

[34] M. Gabbrielli, S. Giallorenzo, C. Guidi, J. Mauro, en F. Montesi, "Self-reconfiguring microservices", in Theory and Practice of Formal Methods, Springer, 2016, bll 194–210.

[35] W. Scarborough, C. Arnold, en M. Dahan, "Case study: Microservice evolution and software lifecycle of the xsede user portal api", in Proceedings of the XSEDE16 Conference on Diversity, Big Data, and Science at Scale, 2016, bll 1–5.

[36] D. Jaramillo, D. V. Nguyen, en R. Smart, "Leveraging microservices architecture by using Docker technology", in SoutheastCon, 2016, bll .5–1.

[37] P. O'Neill en A. S. Sohal, "Business Process Reengineering A review of recent literature", Technovation, vol 19, no 9, bll 571–581, 1999.

[38] G. Toffetti, S. Brunner, M. Blöchlinger, F. Dudouet, en A. Edmonds, "An architecture for self-managing microservices", in Proceedings of the 1st International Workshop on Automated Incident Management in Cloud, 2015, bll 19–24.

[39] A. Messina, R. Rizzo, P. Storniolo, en A. Urso, "A simplified database pattern for the microservice architecture", in The Eighth International Conference on Advances in Databases, Knowledge, and Data Applications (DBKDA), 2016, bll 35–40.

[40] V. D. Le et al., "Microservice-based architecture for the NRDC", in 2015 IEEE 13th International Conference on Industrial Informatics (INDIN), 2015, bll 1659–1664.

[41] J. Stubbs, W. Moreira, en R. Dooley, "Distributed systems of microservices using docker and serfnode", in 2015 7th International Workshop on Science Gateways, 2015, bll 34–39.

[42] M. Abdelbaky, J. Diaz-Montes, M. Parashar, M. Unuvar, en M. Steinder, "Docker containers across multiple clouds and data centers", in 2015 IEEE/ACM 8th International Conference on Utility and Cloud Computing (UCC), 2015, bll 368–371.

[43] A. Messina, R. Rizzo, P. Storniolo, en A. Urso, "A simplified database pattern for the microservice architecture", in The Eighth International Conference on Advances in Databases, Knowledge, and Data Applications (DBKDA), 2016, bll 35–40.

[44] A. Messina, R. Rizzo, P. Storniolo, M. Tripiciano, en A. Urso, "The database-is-the-service pattern for microservice architectures", in International Conference on Information Technology in Bio-and Medical Informatics, 2016, bll 223–233.

[45] A. Versteden, E. Pauwels, en A. Papantoniou, "An Ecosystem of User-facing Microservices Supported by Semantic Models", in USEWOD-PROFILES@ESWC, 2015, vol 1362, bll 12–21.

[46] D. Guo, W. Wang, G. Zeng, en Z. Wei, "Microservices architecture based cloudware deployment platform for service computing", in 2016 IEEE Symposium on Service-Oriented System Engineering (SOSE), 2016, bll 358–363.

[47] A. Balalaie, A. Heydarnoori, en P. Jamshidi, "Migrating to cloud-native architectures using microservices: an experience report", in European Conference on Service-Oriented and Cloud Computing, 2015, bll 201–215.

[48] D. Taibi, V. Lenarduzzi, en C. Pahl, "Architectural patterns for microservices: a systematic mapping study", in *CLOSER 2018: Proceedings of the 8th International Conference on Cloud Computing and Services Science; Funchal, Madeira, Portugal*, 19-21 March 2. 2018.

[49] M. Vianden, H. Lichter, en A. Steffens, "Experience on a microservice-based reference architecture for measurement systems", in 2014 21st Asia-Pacific Software Engineering Conference, 2014, vol 1, bll 183–190.

[50] F. Leymann, C. Fehling, S. Wagner, en J. Wettinger, "Native cloud applications: Why virtual machines, images and containers miss", in Proceedings of the 6th International Conference on Cloud Computing, bll 7–15.

[51] C. Gadea, M. Trifan, D. Ionescu, M. Cordea, en B. Ionescu, "A microservices architecture for collaborative document editing enhanced with face recognition", in 2016 IEEE 11th International Symposium on Applied Computational Intelligence and Informatics (SACI), 2016, bll 441 – 446.

[52] T. Vresk en I. Čavrak, "Architecture of an interoperable IoT platform based on microservices", in 2016 39th International Convention on Information and Communication Technology, Electronics and Microelectronics (MIPRO), 2016, bll 1196–1201.

[53] R. Kewley, N. Kester, en J. McDonnell, "DEVS Distributed Modeling Framework-A parallel DEVS implementation via microservices", in 2016 Symposium on Theory of Modeling and Simulation (TMS-DEVS), 2016, bll 1–8.

[54] H.-M. Chen, R. Kazman, S. Haziyev, V. Kropov, en D. Chtchourov, "Architectural support for DevOps in a neo-metropolis BDaaS platform", in 2015 IEEE 34th symposium on reliable distributed systems workshop (SRDSW), 2015, bll 25–30.

[55] J. Stubbs, W. Moreira, en R. Dooley, "Distributed systems of microservices using docker and serfnode", in 2015 7th International Workshop on Science Gateways, 2015, bll 34–39.

[56] D. Savchenko en G. Radchenko, "Microservices validation: Methodology and implementation", in CEUR Workshop Proceedings. Vol. 1513: Proceedings of the 1st Ural Workshop on Parallel, Distributed, and Cloud Computing for Young Scientists (Ural-PDC 2015). Yekaterinburg, 2015.

[57] R. O'Connor, P. Elger, en P. M. Clarke, "Exploring the impact of situational context - a case study of a software development process for a microservices architecture", in 2016 IEEE/ACM International Conference on Software and System Processes (ICSSP), 2016, bll 6–10.

[58] D. Jaramillo, D. V. Nguyen, en R. Smart, "Leveraging microservices architecture by using Docker technology", in SoutheastCon, 2016.

[59] M. Amaral, J. Polo, D. Carrera, I. Mohamed, M. Unuvar, en M. Steinder, "Performance evaluation of microservices architectures using containers", in IEEE 14th International Symposium on Network Computing and Applications, 2015, bll 27–34.